\documentclass[letter,longauth,traditabstract]{aa} 
\usepackage{graphicx}
\usepackage{txfonts}
\begin{document}
   \title{Cold dust around nearby stars (DUNES). First results\thanks{Herschel is an ESA space observatory with science instruments provided by European-led Principal Investigator consortia and with important participation from NASA.}}

   \subtitle{A resolved exo-Kuiper belt around the solar-like star $\zeta^2$ Ret}

    \author{C. Eiroa
            \inst{1} 
\and D. Fedele 
            \inst{1,2,3}
\and J. Maldonado\inst{1}
\and B.M. Gonz\'alez-Garc\'\i a\inst{4}
 \and J. Rodmann\inst{5}
 \and A. M. Heras\inst{6}
 \and G.L. Pilbratt\inst{6}
 \and J.-Ch. Augereau\inst{7}  
 \and A. Mora\inst{8,1}  
 \and B. Montesinos\inst{9}
 \and D. Ardila\inst{10}
 \and G. Bryden\inst{11}
 \and R. Liseau\inst{12}
 \and K. Stapelfeldt\inst{11}
 \and R. Launhardt\inst{2}
 \and E. Solano\inst{9}
 \and A. Bayo\inst{13}
\and O. Absil\inst{14}
\and M. Ar\'evalo\inst{9}
 \and D. Barrado\inst{9}
 \and C. Beichmann\inst{15}
 \and W. Danchi\inst{16}
 \and C. del Burgo\inst{17}
\and S. Ertel\inst{31}
 \and M. Fridlund\inst{6}
 \and M. Fukagawa\inst{18}
\and R. Guti\'errez\inst{9}
 \and E. Gr\"un\inst{19}
 \and I. Kamp\inst{20}
 \and A. Krivov\inst{21}
\and J. Lebreton\inst{7}
 \and  T. L\"ohne\inst{21}
 \and R. Lorente\inst{22}
 \and  J. Marshall\inst{23}
 \and R. Mart\'\i nez-Arn\'aiz\inst{24}
 \and G. Meeus\inst{1}
 \and  D. Montes\inst{24}
 \and A. Morbidelli\inst{25}
\and S. M\"uller\inst{21}
 \and  H. Mutschke\inst{21}
 \and T. Nakagawa\inst{26}
 \and G. Olofsson\inst{27}
 \and I. Ribas\inst{28}
 \and A. Roberge\inst{16}
 \and J. Sanz-Forcada\inst{9}
 \and P. Th\'ebault\inst{29} 
 \and H. Walker\inst{30}
 \and G. J. White\inst{23,30}
 \and S. Wolf\inst{31}
           }
   \institute{Dpt. F\'\i sica Te\'orica, Facultad de Ciencias, 
            Universidad Aut\'onoma de Madrid,
            Cantoblanco, 28049 Madrid, Spain
             \email{carlos.eiroa@uam.es}
        \and Max-Planck Institut f\"ur Astronomie, K\"onigstuhl 17, 69117 Heidelberg,
             Germany
        \and John Hopkins University, Dept. of Physics and Astronomy, 3701 San Martin drive, Baltimore, MD 21210, USA
        \and INSA at ESAC, E-28691 Villanueva de la Ca{\~n}ada, Madrid, Spain
 \and ESA Space Environment and Effects Section, ESTEC, PO Box 299, 2200 AG Noordwijk, 
 The Netherlands
 \and ESA Astrophysics \& Fundamental Physics Missions Division, ESTEC/SRE-SA,  Keplerlaan 1, NL-2201 AZ Noordwijk,
 The Netherlands
 \and Universit\'e Joseph Fourier/CNRS, Laboratoire d'Astrophysique de Grenoble, UMR 5571, Grenoble, France
 \and ESA-ESAC Gaia SOC. P.O. Box 78
 E-28691 Villanueva de la Ca{\~n}ada, Madrid, Spain
 \and LAEX, Centro de Astrobiolog\'\i a (INTA-CSIC), LAEFF Campus, European Space Astronomy Center (ESAC), P.O.Box 78, E-28691 Villanueva de la Ca{\~n}ada, Madrid, Spain
 \and NASA Herschel Science Center, California Institute of Technology, 1200 E. California Blvd., Pasadena, CA 91125, USA
 \and Jet Propulsion Laboratory, California Institute of Technology, Pasadena, CA 91109, USA
 \and Onsala Space Observatory, Chalmers University of Technology, Se-439 92 Onsala, Sweden
 \and European Space Observatory, Alonso de Cordova 3107, Vitacura
Casilla 19001, Santiago 19, Chile 
 \and Institut d'Astrophysique et de G{\'e}ophysique, Universit{\'e} de Li{\`e}ge, 17 All{\'e}e du Six Ao{\^u}t, B-4000 Sart Tilman, Belgium 
 \and NASA ExoPlanet Science Institute California Inst. of Technology, 1200 E. California Blvd., Pasadena, CA 91125, USA 
 \and NASA Goddard Space Flight Center, Exoplanets and Stellar Astrophysics, Code 667, Greenbelt, MD 20771.USA
 \and UNINOVA-CA3, Campus da Caparica, Quinta da Torre, Monte de Caparica,
2825-149 Caparica, Portugal 
 \and Nagoya University, Japan. 
 \and Max-Planck-Institut f\"ur Kernphysik, Saupfercheckweg 1, D-69117 Heidelberg, Germany 
  \and Kapteyn Astronomical Institute, Postbus 800, 9700 AV Groningen, The Netherlands
  \and   Astrophysikalisches Institut und Universit{\"a}tssternwarte,   
  Friedrich-Schiller-Universit{\"a}t,
  Schillerg{\"a}{\ss}chen 2-3, 07745 Jena, Germany
 \and Herschel Science Center, ESAC/ESA, P.O. BOX 78, 28691 Villanueva de la 
Ca\~nada, Madrid, Spain
 \and  Department of Physics and Astrophysics, Open University, Walton Hall, Milton Keynes MK7 6AA, UK
 \and Universidad Complutense de Madrid, Facultad de Ciencias F\'\i sicas, Dpt. Astrof\'\i sica, av. Complutense s/n. 28040 Madrid, Spain
 \and Observatoire de la C\^ote d'Azur, Boulevard de l'Observatoire, B.P. 4229, 06304 Nice Cedex 4, France. 
 \and Institute of Space and Astronautical Science (ISAS), Japan Aerospace Exploration Agency (JAXA), 3-1-1, Yoshinodai, Sagamihara, Kanagawa, 229-8510, Japan 
 \and Department of Astronomy, Stockholm University, AlbaNova University Center, Roslagstullsbacken 21, SE-106 91 Stockholm, Sweden    
 \and Institut de Ci\`encies de l'Espai (CSIC-IEEC), Campus UAB,
Facultat de Ci\`encies, Torre C5, parell, 2a pl., E-08193 Bellaterra, Barcelona, Spain     
 \and LESIA, Observatoire de Paris, 92195 Meudon France  
 \and Rutherford Appleton Laboratory, Chilton OX11 0QX, UK  
  \and Christian-Albrechts-Universit\"at zu Kiel, Institut f\"ur Theoretische Physik und Astrophysik, Leibnizstr. 15, 24098 Kiel, Germany 
            }


%
\date{Accepted May 11, 2010}

 
  \abstract
{We present the first far-IR observations of the solar-type stars $\delta$ Pav, HR 8501, 51 Peg and $\zeta^2$ Ret, 
taken within the context of the DUNES {\it Herschel} Open Time
  Key  Programme (OTKP). This project  
 uses the PACS and SPIRE instruments with the objective of studying infrared excesses due to exo-Kuiper belts around nearby solar-type stars. The observed 100 $\mu$m fluxes from $\delta$ Pav, HR 8501, and 51 Peg agree with the predicted photospheric fluxes, excluding debris disks  brighter than $L_{\rm dust}/L_\star \sim 5 \times 10^{-7}$ (1 $\sigma$ level) around those stars. A flattened, disk-like structure with a semi-major axis of $\sim$ 100 AU in size is detected around $\zeta^2$ Ret. The resolved structure suggests the presence of an eccentric dust ring, which we interpret as an exo-Kuiper belt with 
  $L_{\rm dust}/L_\star$ $\approx$ 10$^{-5}$.}

   \maketitle
%

\section{Introduction}

The discovery of infrared  excess emission produced by cold, optically
thin disks composed of  micron-sized dust grains around main
sequence stars is one of the main contributions of IRAS (Aumann et al.
\cite{aumann}). Since the lifetime of  such grains, set by destructive collisions,
Poynting-Robertson  drag and radiation pressure, is much  shorter than
the ages  of these stars,  one must conclude  that those dust  disks -
called debris  disks - are  continuously replenished by  collisions of
large rocky bodies (Backman \& Paresce \cite{backman}).
Observations  of debris  disks provide  powerful  diagnostics
  from  which  to  learn  about  the dust  content,  its  properties
and its spatial  distribution; in addition,
since dust sensitively  responds to the gravity of  planets, it can be
used as  a tracer of the  presence of planets.   Thus, observations of
debris disks around stars of different  masses and ages inform  us about the
formation and evolution of planetary  systems, since they are a direct
proof  for the  existence of  planetesimals and  an indirect 
  tracer of the presence of planets around stars.

  In the Solar  System, the asteroid and Kuiper  belts are examples of
  debris   systems;  in particular, the Kuiper belt has an estimated   dust   
luminosity   $L_{\rm dust}/L_\star \sim  10^{-7}$-$10^{-6}$ (Stern \cite{stern}).
IRAS was  only able to detect bright disks,
  $L_{\rm dust}/L_\star >  10^{-4}$, mainly around A and  F stars; ISO
  extended our  knowledge to a wider  spectral type range  and found a
  general decline  with the stellar  age (Habing et  al \cite{habing},
  Decin  et al.  \cite{decin}).  A  remarkable step  forward  has been
  achieved  by {\it Spitzer},  studying  debris disks  as  faint as  $L_{\rm
    dust}/L_\star$ several times $10^{-6}$,  their incidence from A to
  M type  stars, the age  distribution, the presence of  planets, etc.
  (e.g.  Su  et al.  \cite{su},  Trilling  et al.   \cite{trilling08},
  Bryden  et  al.  \cite{bryden09}).   {\it Spitzer}  has, however,  several
  limitations.   Its   poor  spatial   resolution   prevent  us
from constraining  fundamental disk
parameters which require resolved imaging,    
and the confusion limit inherent in its large beam  limits its
detection capability  to cold disks  brighter than the Kuiper  belt by
around  two  orders of  magnitude.  
Also, {\it Spitzer} is not sensitive longward of 70 $\mu$m,
wavelengths particularly important for the cold disks generally 
found around Sun-like stars.
  The far-infrared 3.5 m diameter {\it Herschel} space telescope
  (Pilbratt  et  al.    \cite{pilbratt})  with  its  instruments  PACS
  (Poglistsch  et al.   \cite{poglitsch})  and SPIRE  (Griffin et  al.
  \cite{griffin})   overcomes    these   limitations,   offering   the
  possibility  of characterising cold,  $\sim$ 30  K, debris  disks as
  faint  as $L_{\rm  dust}/L_\star$ few  times $10^{-7}$  with spatial
  resolution  $\sim$ 30  AU at  10 pc,  i.e., true  extra-solar Kuiper
  belts.

DUNES\footnote{{\em   DU}st   around   {\em   NE}arby   {\em   S}tars,
  http://www.mpia-hd.mpg.de/DUNES/.}  is a {\it Herschel} OTKP  designed  to detect  and characterize  cold,
faint, debris disks,  i.e., extra-solar analogues to the Kuiper belt, around a
statistically  significant sample of  main-sequence FGK  nearby stars,
taking advantage of the unique  capabilities of {\it Herschel} with PACS and
SPIRE.  
The data will  be analysed with  
radiative,
collisional and dynamical dust disk models. 
A complete description of
DUNES goals,  target selection, and  stellar properties
will  be presented  elsewhere (Eiroa  et al.   2010,  in preparation).
The objectives of the DUNES survey are complementary to those
of the  OTKP DEBRIS (Matthews et al.   \cite{matthews}). Both projects
have complementary  star samples, sharing partly some  sources and the
corresponding data.
 
 The DUNES  objectives  require  the detection of  very faint excesses  
at the mJy 
level, comparable  to  the   photospheric  emission 
and only a few times the measurement uncertainties. The
primary observing  strategy is designed  to integrate  for as
  long as needed  to detect the 100 $\mu$m  photospheric flux, subject
  only to confusion noise limitations.  In this letter  we present  our first
results   obtained   during  the   science   demonstration  phase (SDP)  of
{\it Herschel}. 
Four solar-type G stars were  observed:  $\zeta^2$  Ret,  $\delta$  Pav,  
HR 8501, and 51  Peg.  We also observed q$^1$ Eri
as  a test object with a well-known, bright
debris  disk;  the  q$^1$   Eri  results  are  presented  in  an
accompanying letter (Liseau et al. \cite{liseau}). Excluding  $\delta$  Pav,
the rest of the stars are shared targets with DEBRIS. 

\section {Observations and data reduction}
\begin{table}
\label{AORs}
\caption{Summary of the SDP DUNES observations.}
\begin{tabular}{lclrcr}
\hline
Star          & Obs. ID   &Mode\tablefootmark{1}& 
\multicolumn{1}{c}{Bands} & Scan & 
\multicolumn{1}{c}{OT} \\
              &           &    & \multicolumn{1}{c}{($\mu$m)} & Direction    & (sec)     \\
\hline
q$^1$ Eri     & 1342187139/40& SM & 100/160 &63$\degr$/117$\degr$    &2536       \\
q$^1$ Eri     & 1342187141& PS & 100/160 &              &4714       \\
q$^1$ Eri     & 1342187142& PS & 70/160  &              &789        \\
$\zeta^2$ Ret & 1342183660& PS & 100/160 &              &1572       \\
$\zeta^2$ Ret & 1342191102/03& SM\tablefootmark{2} & 100/160 &117$\degr$/63$\degr$       &4510       \\
$\zeta^2$ Ret & 1342191104/05& SM\tablefootmark{2} & 70/160  &63$\degr$/17$\degr$ &4510       \\
$\delta$  Pav & 1342187075/06& SM & 100/160 &45$\degr$/135$\degr$    &3834       \\
HR 8501     & 1342187145/46& SM & 100/160 &63$\degr$/117$\degr$    &1890        \\
51 Peg        & 1342187255& PS & 100/160 &              &1731       \\
\hline
\end{tabular}
\tablefoottext{1}{SM = scan map; PS = chop-nod/point-source mode}

\tablefoottext{2}{Routine phase DUNES observing time (not SDP)}
\end{table}%

The  stars were observed  with PACS  at 70  $\mu$m (blue),  100 $\mu$m
(green), and 160 $\mu$m (red). 
Two observing modes were used - chop-nod/point-source (PS hereafter) and scan map (SM hereafter).
Our data set allows
us to make  a direct comparison of both modes,  specially in the cases
of  q$^1$  Eri  and  $\zeta^2$  Ret. A  critical  evaluation  of  this
comparison will be the subject of a technical note. SM  observations of 
$\zeta^2$ Ret  were carried out as DUNES routine  observations.
Table  1  gives  some   details  of  the  observations  including  the
identification number,  the observing  mode, the 
wavelength bands, 
the scan
direction angles, 
and the total duration of the observation (OT).

Data  reduction   was  carried  out   using  the  {\it Herschel}
Interactive Processing Environment (HIPE), version v2.0.0 RC3, and the
pipeline script delivered at the December 2009 {\it Herschel} data reduction
workshop,  ESAC, Madrid,  Spain.  The script  provides  all the  tools
to  convert pure  raw PACS/{\it Herschel} data  to flux calibrated and position rectified images.  While the instrument pixel size is 3$\farcs$2 for the blue and green 
bands and 6$\farcs$4 for the red band,
the resolution of the final images is set to 1$\arcsec$/pixel and 
2$\arcsec$/pixel for the blue/green and red bands, respectively.

\section{Results}

\subsection{$\delta$ Pav, HR 8501, 51 Peg}

Table  2 gives  J2000.0  equatorial coordinates  of  $\delta$ Pav,  HR 8501 and 51  Peg at 100 $\mu$m, as well as  their optical positions. PACS 
positions are corrected   from  the proper  motions   of  the  stars.
Differences between  the optical and  100 $\mu$m positions  are within
the uncertainties for  {\it Herschel} pointing. Of the three stars,
only $\delta$  Pav has  been detected at  160 $\mu$m.  

\begin{table*}[ht]
\label{pointing}
\caption{Equatorial coordinates, FWHM at 100 $\mu$m, 
observed fluxes with 1$\sigma$ statistical errors ($F_{\rm PACS}$), 
and predicted photospheric fluxes ($F_{\star}$). 
Absolute PACS uncertainties are $\sim$10\% and less than 20\% for 100 and 160 $\mu$m, respectively. 
Flux units are mJy.}
\begin{center}
\begin{tabular}{lccccrrrr}
\hline
Star          & Optical position & PACS 100 $\mu$m position
& \multicolumn{3}{c}{PACS 100 $\mu$m} 
& \multicolumn{2}{c}{PACS 160 $\mu$m} \\
        & (J2000)      & (J2000) &FWHM 
& $F_{\rm PACS}$ & $F_{\star}$
& $F_{\rm PACS}$ & $F_{\star}$ \\
\hline                                                                          $\delta$ Pav&20:08:43.61 -66:10:55.4 &20:08:43.53 -66:10:58.1&6$\farcs$3$\times$6$\farcs$2&59.6$\pm$1.1   &68.7&21.0$\pm$1.5   &26.9 \\
HR 8501     &22:18:15.62 -53:37:37.5 &22:18:16.14 -53:37:37.1&6$\farcs$3$\times$6$\farcs$2& 9.8$\pm$1.2   &10.9&       & 4.3 \\
51 Peg      &22:57:27.98 +20:46:07.8 &22:57:28.44 +20:46:08.5&5$\farcs$9$\times$6$\farcs$0&11.3$\pm$1.7   &10.8&       & 4.2 \\
\hline
\end{tabular}
\end{center}

\end{table*}%

\begin{figure*}
   \centering
      \includegraphics[width=13.6cm]{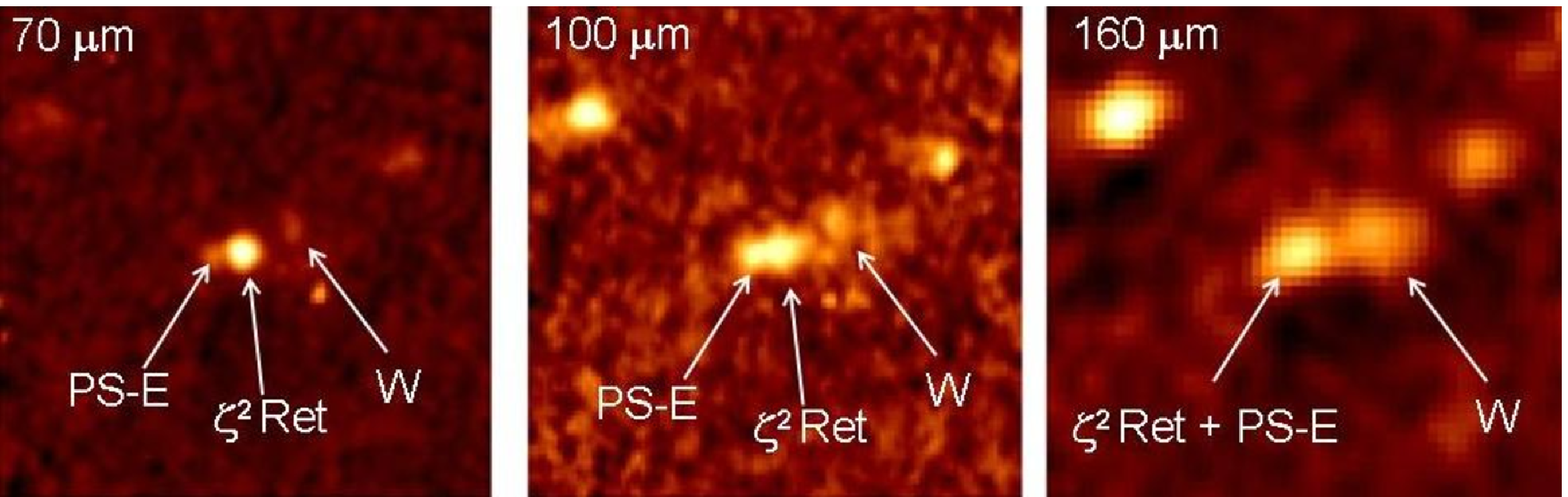}
   \includegraphics[width=3.8cm,angle=-90]{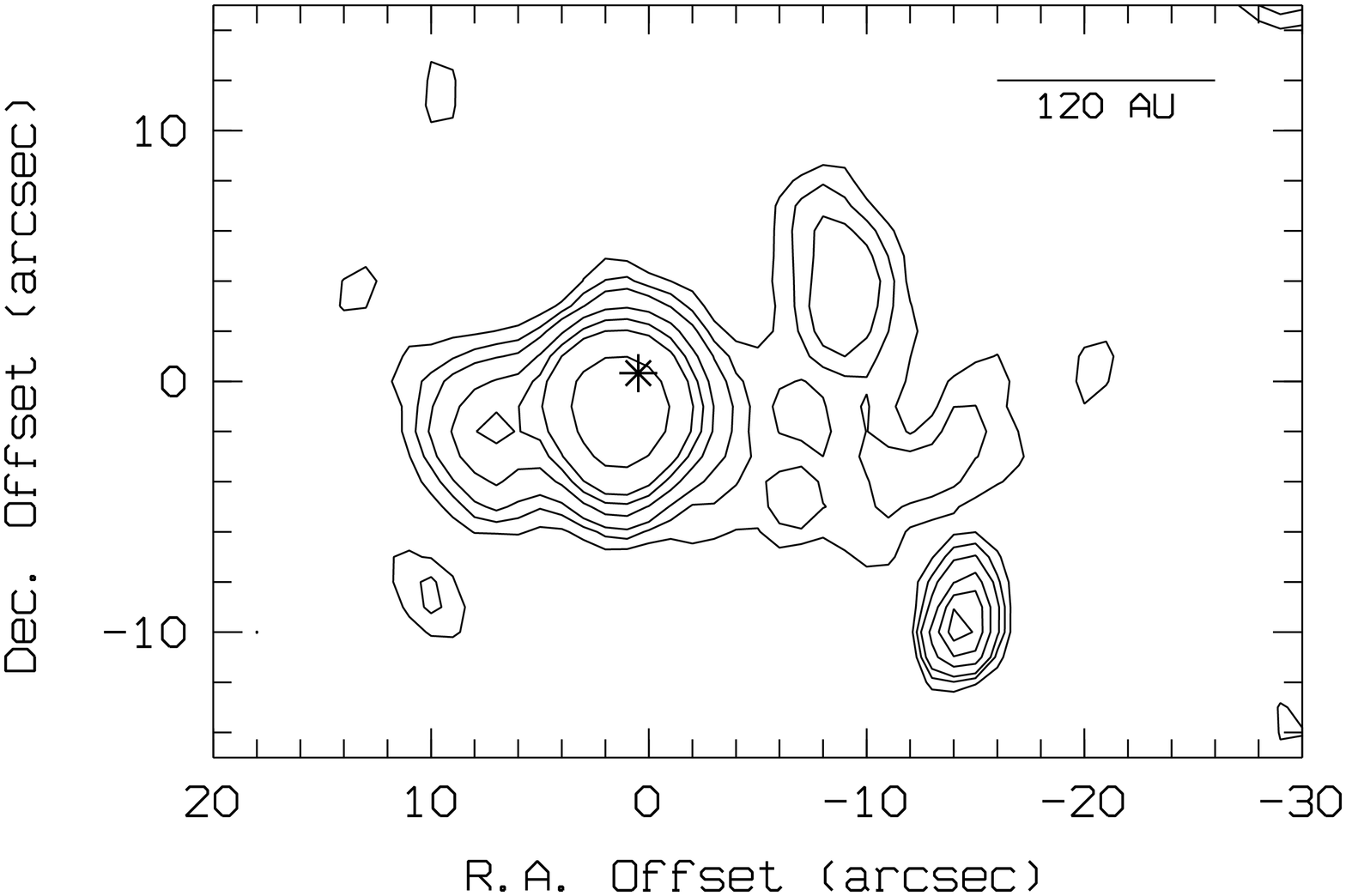}
   \includegraphics[width=3.8cm,angle=-90]{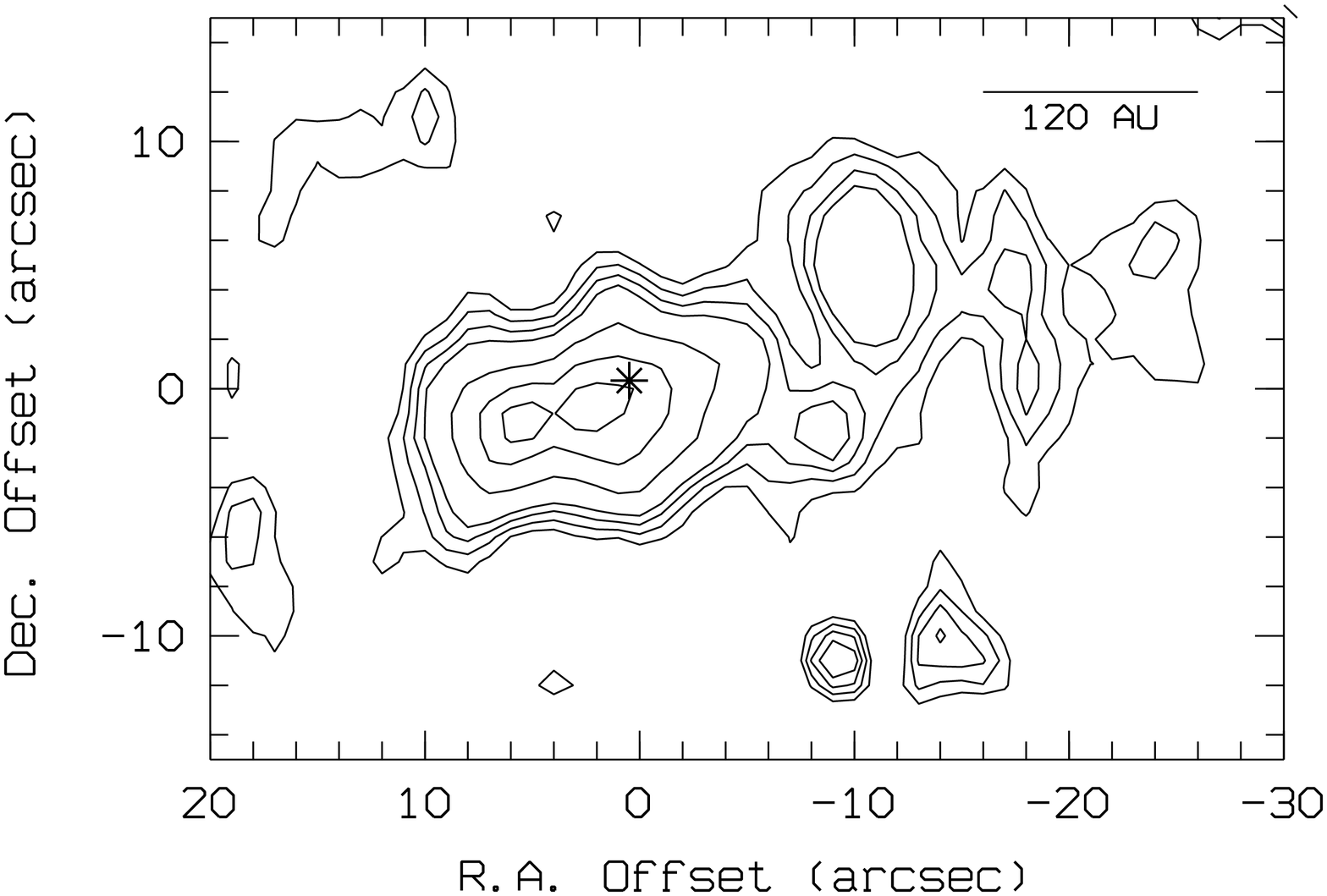}
   \includegraphics[width=3.8cm,angle=-90]{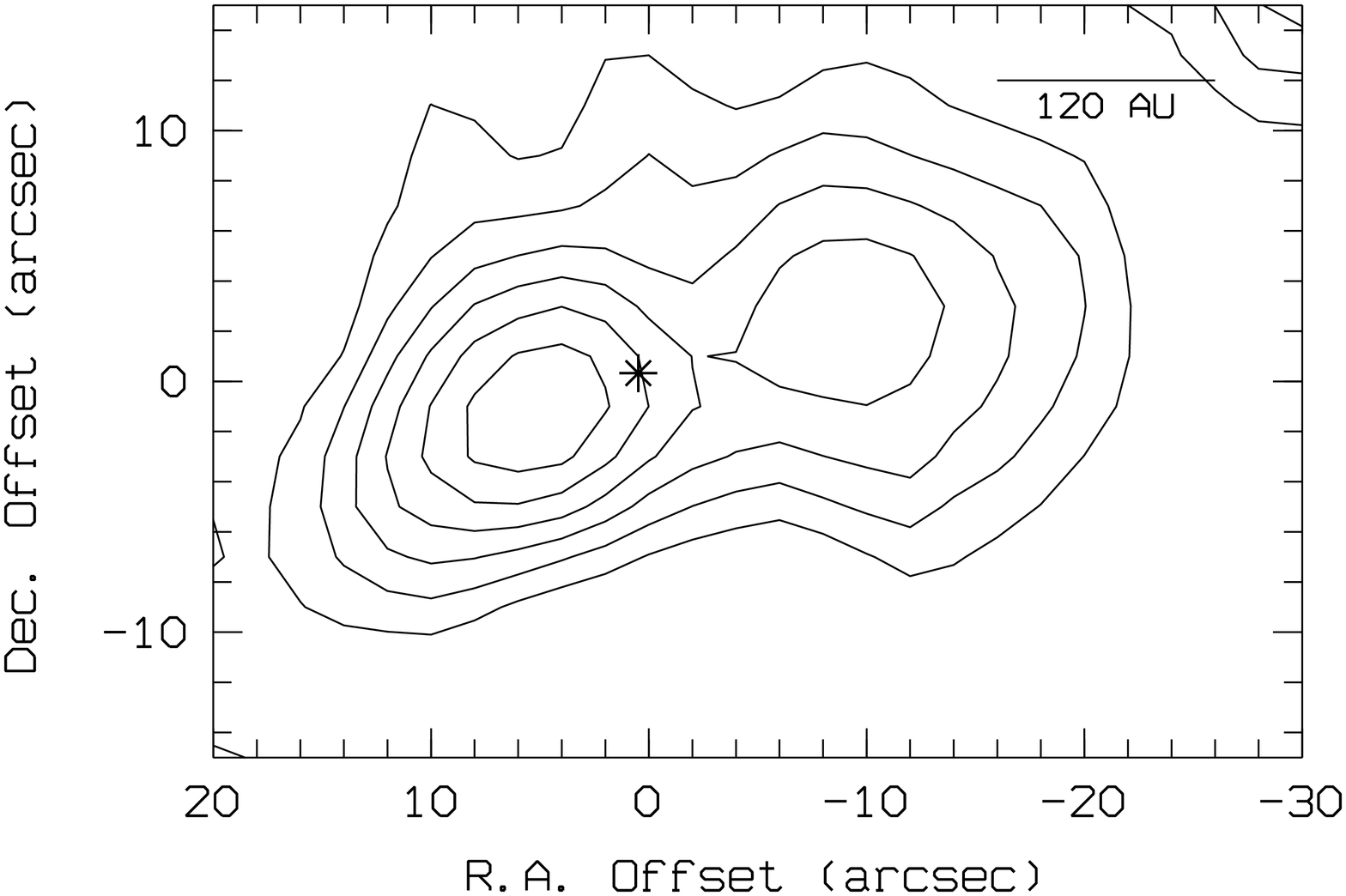}
   \includegraphics[width=3.8cm,angle=-90]{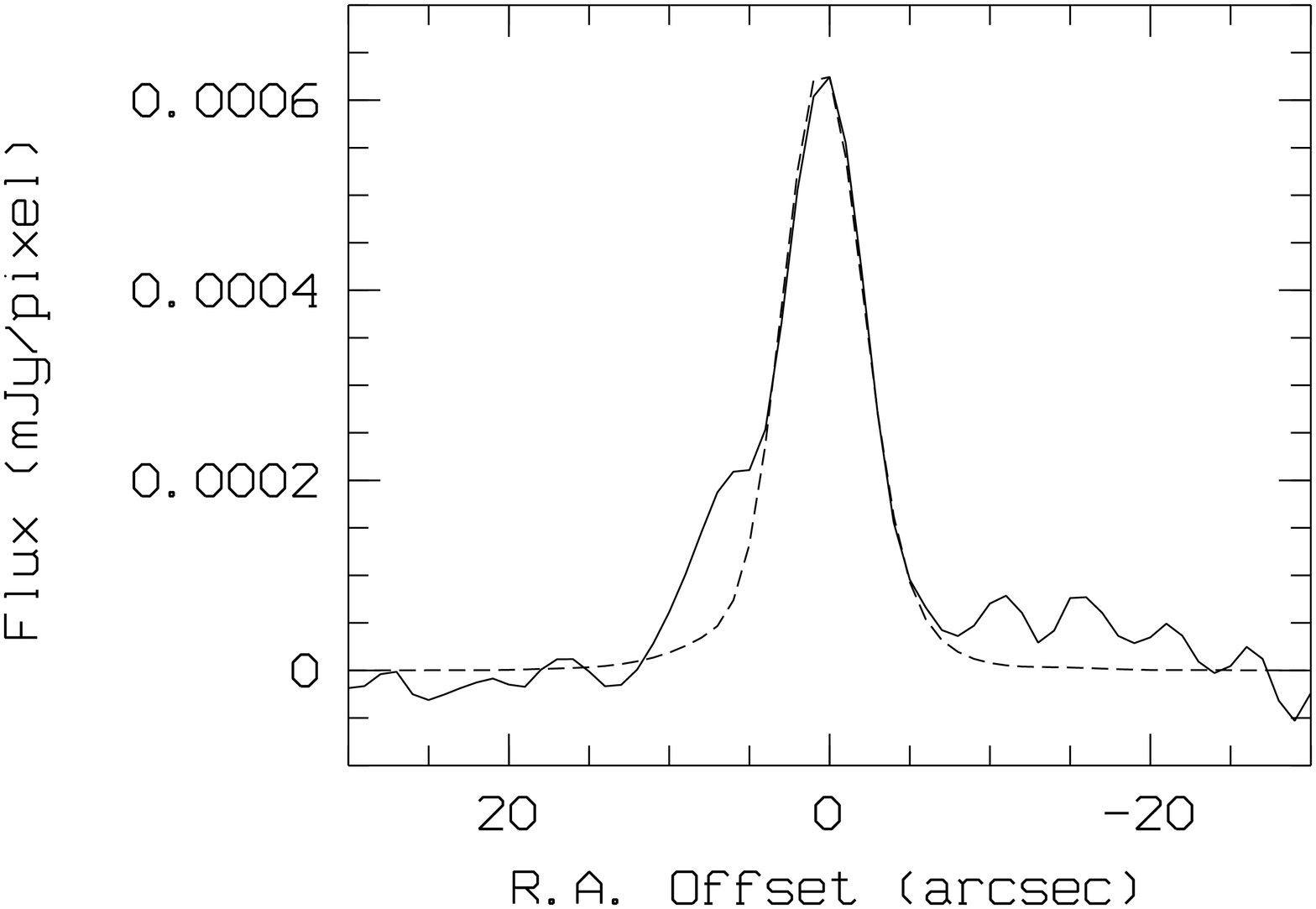}
   \includegraphics[width=3.8cm,angle=-90]{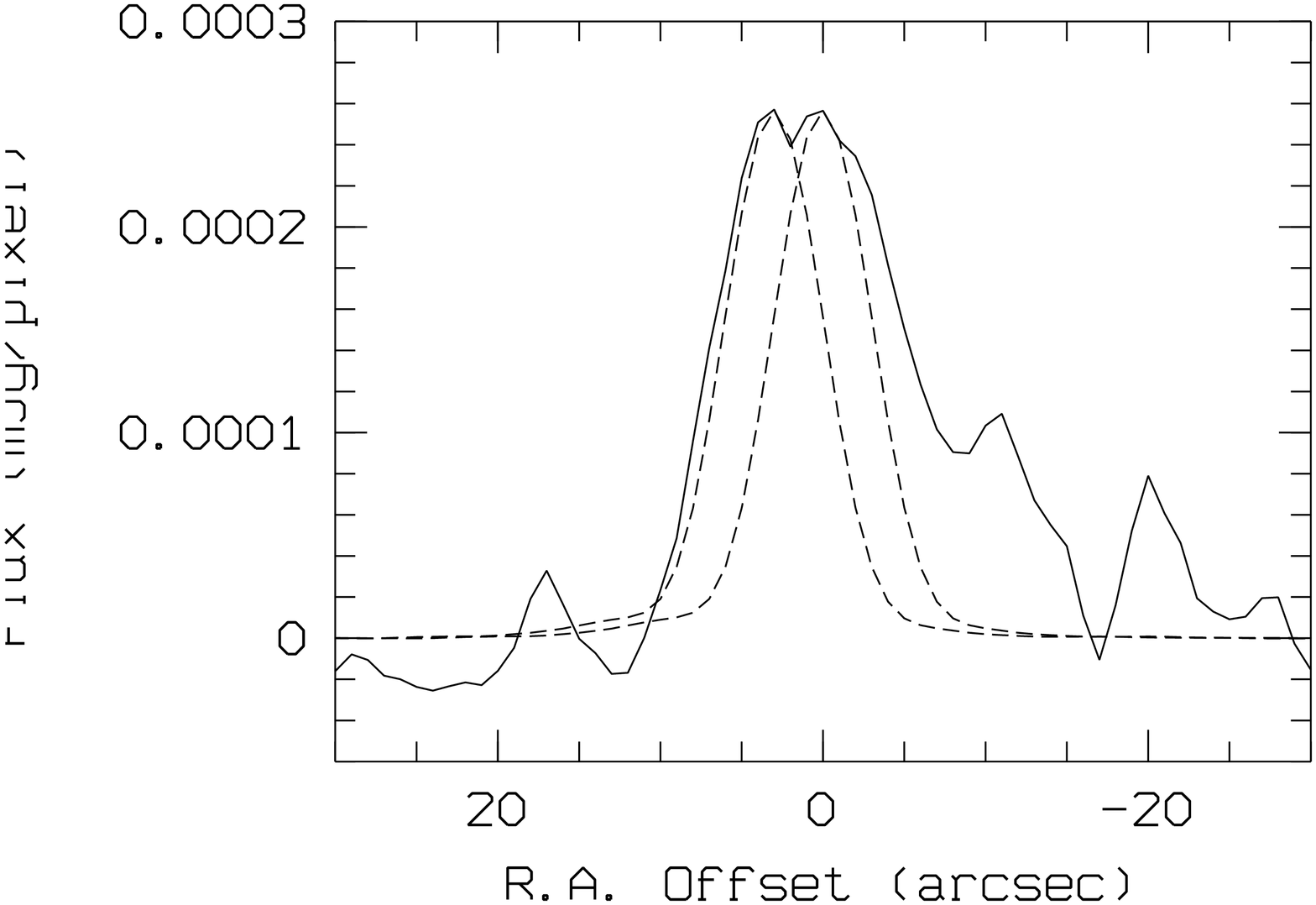}
   \includegraphics[width=3.8cm,angle=-90]{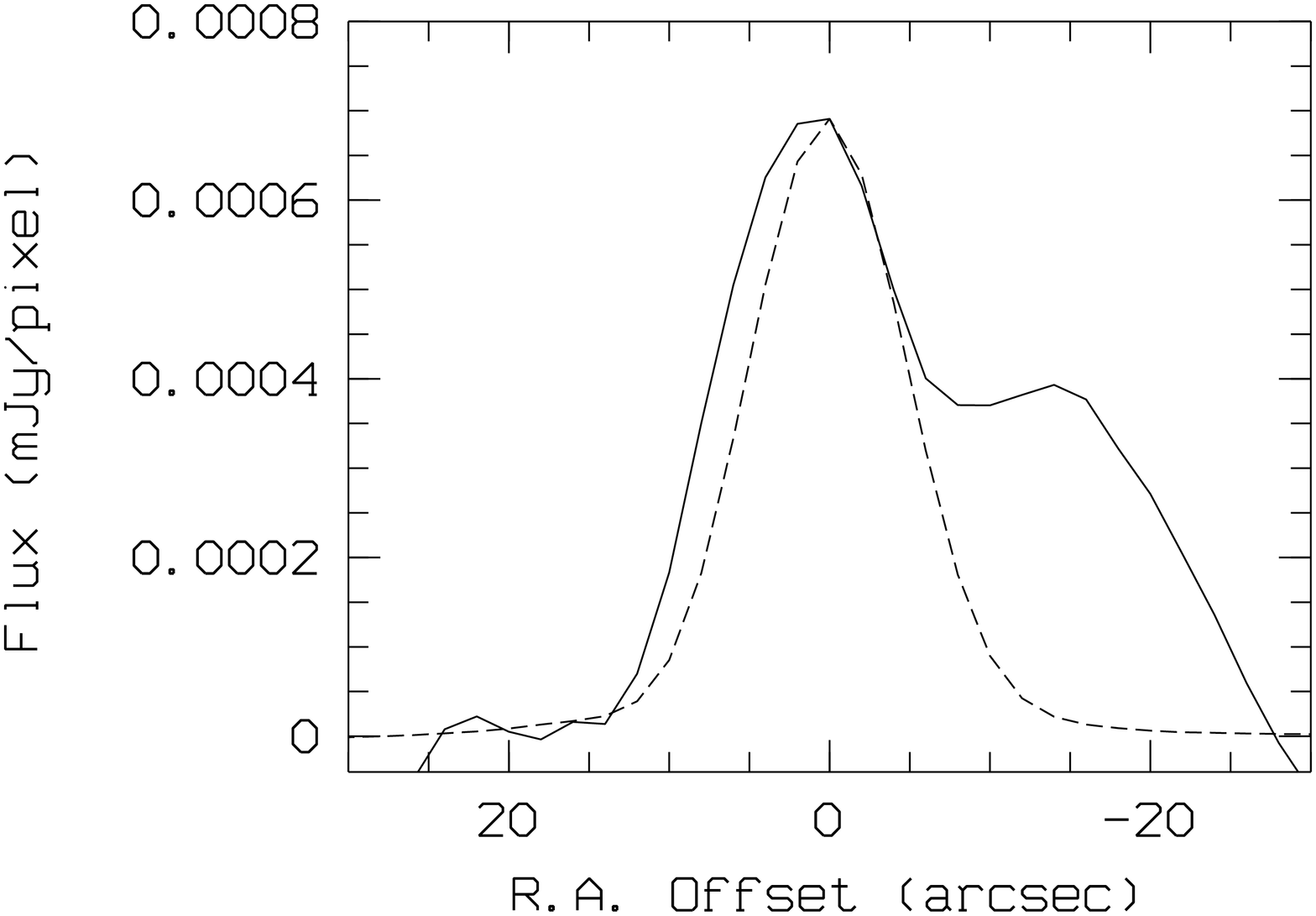}

\caption{PACS results of  $\zeta^2$ Ret. Panels from left to right:  
70 $\mu$m, 100 $\mu$m, and 160 $\mu$m.
Upper panel: Field size  is $100\arcsec\times100\arcsec$ with   East to the left and North
up. Middle  panels plot isocontours. Note that the  field size is different  
from that displayed  in the upper panels. The  ``star'' symbol, position (0,0), corresponds to the optical star (Table 3). A segment 
indicating 120 AU is shown. Contours: 70  $\mu$m:  3,4,6,9,12,15,24 $\sigma$,  
100 $\mu$m: 3,4,5,6,12,14 $\sigma$,  160  $\mu$m:  3,6,9,12,15,18 $\sigma$ ( 
1$\sigma$ values are indicated in the text). Lower panels:  R.A.  intensity  profiles 
through the  peak  flux of $\zeta^2$  Ret (solid line). For  
comparison,  R.A. intensity profiles of  the calibration star  $\alpha$ Bootis 
are  also plotted (dashed lines), scaled  to the peak flux  of  $\zeta^2$ Ret (70 and
  100 $\mu$m) and $\zeta^2$ Ret+PS-E (160 $\mu$m). The $\alpha$ Bootis 
profile is also scaled and superimposed to PS-E in the 100 $\mu$m panel. }
    \end{figure*}

100 $\mu$m FWHM values of $\delta$ Pav have been estimated using a
2-D  gaussian  fit. This  procedure did  not  produce  reasonable
results for HR 8501, perhaps  due to the faintness of the star;
in this case  the FWHM has been estimated  from intensity profiles
along the R.A.  and Dec.  directions.  For 51  Peg, observed in PS
mode,  the  FWHM  estimate   is  also  based  on  some  additional
point-like sources visible in  the PACS field. The 100 $\mu$m FWHM estimates 
are given in  Table 2.  The 160  $\mu$m 2-D gaussian  fit for $\delta$
Pav gives FWHM = 11$\farcs$8$\times$9$\farcs$2, with conservative errors 
$\sim 1\arcsec$.  These values and the elongated beam  are
consistent with the expected results for point sources (see the technical notes PICC-ME-TN-035/036
 in http://herschel.esac.esa.int/AOTsReleaseStatus.shtml)
 
Aperture photometry has been used to estimate the flux of the
stars. Table 2 gives  fluxes and  errors taking into 
account the correction 
factors indicated in the aforementioned technical notes. 
 The sky  noise was 2.5$\times 10^{-5}$ Jy/pixel
and 2.7$\times  10^{-5}$ Jy/pixel for the 100  $\mu$m SM data of
$\delta$ Pav  and HR 8501, respectively. 
The sky noise is considerably higher in the PS-mode 51 Peg image
($\approx$$4.3 \times 10^{-5}$ Jy/arcsec$^2$)
due to the very irregular background  and presence of negative signals.
The 160 $\mu$m sky noise was 4.9$\times  10^{-5}$ Jy/pixel 
for the $\delta$ Pav data. The
absolute  calibration uncertainties are  10\% in the
blue  and green  bands and  better  than 20\%  in the  red band.

\subsection{$\zeta^2$ Ret (HIP 15371)}

Fig  1  shows the  SM PACS images of  $\zeta^2$  Ret.  An  East-West
oriented structure  is seen  at 70  and 100  $\mu$m.  It
consists  of two  point-like flux  peaks  embedded in  a  faint, extended
emission, which displays a secondary diffuse  maximum at its Western side. 
Both point-like peaks have similar brightness in the green band,
but the Eastern point-like peak  is much fainter in the blue band. The two 
point-like sources are unresolved in the 
lower-resolution 
160 $\mu$m image; 
instead  a single bright peak
is observed at that position with a secondary maximum at the 
position of the 70/100 $\mu$m  Western diffuse emission.
  
Table  3  gives positions  at  70 and  100  $\mu$m of  both
point-like sources, and of the  brightest 160 $\mu$m peak; the optical
position  of $\zeta^2$  Ret is  also  given for comparison. The  brightest 70  $\mu$m
peak coincides  with  the optical  position  of the star within  the {\it Herschel}  pointing
error; this result and the fact that its PACS 70 and 100 $\mu$m fluxes
are  similar  to the  expected  $\zeta^2$ Ret  photospheric
fluxes (see  below) lead  us to identify  this point-like  PACS object
with the optical  star. There is a small shift between  the 70 and 100
$\mu$m positions of $\zeta^2$ Ret, but we note that a similar shift is
found  for other  field objects  -  a blue  object very  close to  the
$\zeta^2$  Ret complex  towards the  South-West; and  two red  objects, one
towards the Northwest and one towards the Northeast (see Fig. 1).

The middle  panels of  Fig. 1  show isocontour plots. The
optical  position of  $\zeta^2$ Ret  is  marked.  100  $\mu$m  and  
160  $\mu$m contours have been spatially shifted so that the peak positions 
of the mentioned field  objects coincide  in all  three bands 
(those objects are not shown in the isocontour plots).   
The size of 
the 
whole structure  changes with wavelength from  $\approx 25\arcsec  
\times 15\arcsec$  in the  blue  to $\approx 40\arcsec  \times 15\arcsec$  
in  the red  band.  East-West  intensity profiles are  shown in  the 
bottom of  Fig. 1, together  with similar profiles of  $\alpha$ Bootis. The
blue and green  intensity profiles show the point-like character of 
$\zeta^2$ Ret, as well as of the faint peak at the East, called PS-E hereafter 
; the profile in the red band also shows  point-like behaviour for the brightest 160 $\mu$m peak, $\zeta^2$ Ret+PS-E in Fig. 1 and Table 3.  
The Western diffuse peak (``W'' in Fig. 1) appears very prominent in 
the green and red profiles, while  it is  very  faint  compared  to  
$\zeta^2$  Ret in the  blue profile.  North-South profiles  (not  shown)  do 
not  resolved either $\zeta^2$ Ret, PS-E, or $\zeta^2$ Ret+PS-E in  
any band,  i.e.  they are point-like along that  direction
  with no hint of  any faint extended emission.

Table 3 gives PACS fluxes estimated using the flux peaks of the
point-like sources and integrating over beam  sizes given by  
$\pi$ (FWHM$_x$ $\times$  FWHM$_y$) /4 $\ln  2$.  The PACS source identified 
with $\zeta^2$ Ret is a 
blue  object, while PS-E is a  red one.  The flux  at 160
$\mu$m corresponds to   $\zeta^2$ Ret+PS-E, but  
PS-E is the main  contributor to the flux at this wavelength - 
the emission is peaking more  towards  PS-E (Fig. 1).  The total flux of  the $\zeta^2$ Ret complex is  44.5 mJy, 40.4  mJy, and
42.6 mJy in the blue, green and red bands, respectively.  
The estimated 70  $\mu$m flux for the whole complex agrees very well  
with the unresolved {\it Spitzer} flux of 46 mJy at the same wavelength (Trilling et al. \cite{trilling08}).
   
\begin{table}
\caption{Coordinates,fluxes, and 1$\sigma$ statistical errors of the $\zeta^2$ Ret  PACS point-like sources. Flux units are mJy.
}             
\label{table:1}      
\centering          
\begin{tabular}{lccr }     
\hline\hline       
Object                    & $\alpha$(2000.0)  & $\delta$(2000.0)  & Flux    \\ 
\hline 
$\zeta^2$ Ret (optical)       & 03:18:12.82       &-62:30:22.9       &     \\
\hline 
$\zeta^2$ Ret (70 $\mu$m)   & 03:18:12.77       &-62:30:23.2       & 24.9$\pm$0.8\\
$\zeta^2$ Ret (100 $\mu$m)  & 03:18:13.12       &-62:30:24.4       & 13.4$\pm$1.0\\
\hline 
PS-E (70 $\mu$m)   & 03:18:13.63       &-62:30:24.2       & 8.9$\pm$0.8\\
PS-E (100 $\mu$m)  & 03:18:13.55       &-62:30:24.4       & 13.5$\pm$1.0\\
\hline
$\zeta^2$ Ret+PS-E (160 $\mu$m) & 03:18:13.37       &-62:30:23.2       & 19.4$\pm$1.5\\
\hline
\end{tabular}
\end{table}

\section{Discussion}

\begin{figure}
   \centering
\includegraphics[width=7.8cm]{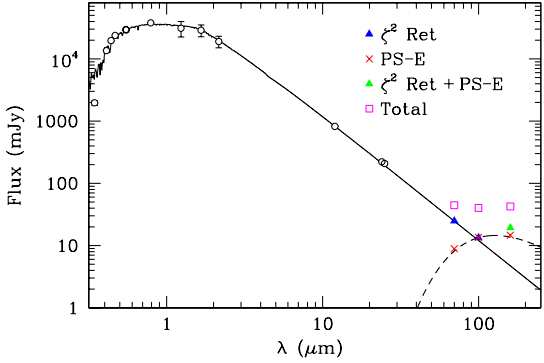}
   \caption{SED of $\zeta^2$ Ret. Optical, 2MASS, IRAS, and {\it Spitzer} fluxes 
are indicated by black symbols. Blue triangles are $\zeta^2$ Ret; 
red crosses are PS-E; green triangle is $\zeta^2$ Ret+PS-E; magenta squares are total fluxes from the $\zeta^2$ Ret complex. Error bars are smaller than the size of the symbols. The solid line is the best $\chi^2$ photospheric fit
(T$_{eff}$ =  5850 K, $\log g$ = 4.5,  and [Fe/H] = -0.23, which are mean values found using 
the DUNES discovery  tool, http://sdc.laeff.inta.es/dunes).
The dashed line is a 40 K black body normalized 
at the PS-E 100 $\mu$m flux. The deduced 160 $\mu$m flux from PS-E is also plotted with a red cross (see text).}         
    \end{figure}

Our data do not reveal any cold dust disk around $\delta$ 
Pav, HR 8501 or 51 Peg, since the observed and predicted 100 $\mu$m photospheric fluxes coincide within the uncertainties (Table 2).
Assuming dust temperatures of 40 K (peak blackbody fluxes at $\sim 100 ~\mu$m),
we can exclude debris disks with $L_{\rm dust}/L_\star \gtrsim 5\times10^{-7}$ (1$\sigma$) (Eqn. 4 from Beichmann et al. \cite{beichmann}).

$\zeta^2$ Ret, located at 12 pc, is a {\bf G1} V star with a bolometric luminosity of 0.97 L$_\odot$ and an estimated age of $\sim$ 2-3 Gyr 
(Eiroa et al. 2010, in prep.).
Figure  2  shows  the stellar  SED  as  well as PACS fluxes  
from PS-E and the whole complex; a PHOENIX  stellar photosphere (Brott \& Hauschildt 
\cite{brott}) is also plotted. The agreement
between the observed  70 and 100 $\mu$m fluxes from the blue bright 
point-source and those predicted by  the photospheric fit, 
24.7 mJy and  12.1 mJy at 70 and 100 $\mu$m, respectively,  is excellent.
 This photometry and its positional alignment 
with the stellar position
support our claim that the PACS blue point-like object is
indeed $\zeta^2$ Ret. On the other hand, the  nature  of  the extended structure is intriguing. 
 While coincidental alignments with background objects are common in
IRAS all-sky images, the much higher resolution of {\it Herschel} makes
such juxtapositions unlikely within a targeted survey.
Based on {\it Spitzer} source counts of background galaxies at 70 $\mu$m
(Dole et al. \cite{Dole04}), we find that the probability of chance alignment with a
$\ge$20 mJy source within 10$\arcsec$ is just $10^{-3}$.

The source PS-E is a red object with a black body 
temperature T(70-100 $\mu$m)  $\approx$ 40 K.  We have pointed  out 
that both PS-E and $\zeta^2$ Ret are not resolved  at 160 $\mu$m and
that the flux peak  at this wavelength  is closer to  PS-E. If  we 
subtract from the  measured 160 $\mu$m flux the stellar flux predicted for  $\zeta^2$ Ret (4.7 mJy) and make the  
plausible  assumption that  the residual flux (14.7 mJy) originates in 
PS-E, this 160 $\mu$m flux for PS-E  is again consistent with  
a $\sim$ 40  K blackbody (see Fig. 2). 
PS-E is clearly not stellar; we suggest that it is instead orbiting circumstellar dust.
The contribution of the extended emission to the  total flux can be
estimated  subtracting  the point-like  sources  from  the total  flux
reported  above.    The  residual  flux  mainly   corresponds  to  the
Western diffuse  emission since the point-like sources are not resolved along the North-South direction. In this  case, the remaining flux corresponds to black body  temperatures in the range $\sim$30-40 K, and the  total fractional luminosity
from the entire structure surrounding $\zeta^2$ is $L_{\rm dust}/L_\star \approx  10^{-5}$.

We  have  the interesting scenario  of a {\bf G1}  V star surrounded by optically-thin 30-40 K  emission.
This is the temperature range expected for
black body  dust grains orbiting at distances  $\sim$100 AU from  the star. 
This is consistent with the projected  linear  distances  from
$\zeta^2$ Ret  to PS-E and  to the Western diffuse emission  
of $\sim$70 AU 
and $\sim$120  AU, respectively. The  red  image  
suggests  a  flattened, disk-like structure  with the  star located  
asymmetrically along  the  major axis, while the blue and green images suggest it is ring-like given the 
flux cavity towards the West from the star. 
We interpret the structure in the PACS images as a dust ring 
surrounding $\zeta^2$ Ret. 
We attribute the observed East-West asymmetry to a 
significant disk eccentricity - $e \approx 0.3$.
Similarly, an offset is observed in the Fomalhaut debris disk 
with $e \approx 0.1$ (Stapelfeldt et al. \cite{stapelfeldt04}).
Maintaining a stable eccentric ring requires an external driving
force such as a shepharding planet (Wyatt et al. \cite{wyatt99})
and in the case of Fomalhaut the predicted planet has been 
been directly imaged (Kalas et al. \cite{kalas08}).
The disk asymmetry in the $\zeta^2$ Ret system may likewise
be the signature of an unseen planetary companion.
 While this is an exciting possibility,
 other forces might also produce disk asymmetry.
For example, interaction with the ISM is probably responsible for the
strong asymmetry observed around HD 61005,
since its brightness offset is well aligned with the star's space motion
(Hines et al. \cite{hines07}).
A more profound analysis and detailed modeling of  $\zeta^2$ Ret and the suggested Kuiper belt 
is deferred to a future paper.

\section{Conclusions}

Our results show the capability of {\it Herschel}/PACS to detect and resolve cold dust disks 
with a luminosity close to the solar Kuiper belt, which will allows us to deepen our 
understanding of planetary systems, in particular those associated with mature stars. 
Specifically, our data exclude the existence of cold debris disks with $L_{\rm dust}/L_\star \gtrsim 5\times10^{-7}$ (1$\sigma$) around the solar-type stars $\delta$ Pav, HR 8501 and 51 Peg. On the other hand, the data show that $\zeta^2$ Ret is a good example where cold disks around nearby stars, very much alike the solar Kuiper belt, can be resolved and studied in great detail with the {\it Herschel} space observatory.


\begin{thebibliography}{}
\bibitem[1984]{aumann} Aumann H.H. et al. 1984 \apj ~278, L23
\bibitem[1993]{backman} Backman, D.E., \& Paresce, F. 1993,, in Protostars and Planets III, ed. E.H. Levy \& J.I. Lunine (Tucson: Univ. Arizona Press), 1253
\bibitem[2006]{beichmann} Beichmann et al. 2006, \apj ~652, 1674
\bibitem[2005]{brott} Brott, I. \&
  Hauschildt, P.H. 2005, ESA SP-576, C. Turon,
  K.S. O'Flaherty, M.A.C. Perryman (eds), p.565 
\bibitem[2009]{bryden09} Bryden et al. 2009, \apj ~705, 1226
\bibitem[2004]{Dole04}
Dole, H., {Le Floc'h}, E., P\'erez-Gonz\'alez, P.~G., et~al. 2004, \apjs, 154, 87 
\bibitem[2003]{decin} Decin, G., Dominik, C., Waters, L.B.F.M. \& Waelkens, C. 2003, \apj ~598, 636
  \bibitem[2010]{griffin} Griffin et al. 2010, this volume
\bibitem[2001]{habing} Habing, H.J. et al. 2001, \aap ~365, 545
\bibitem[2007]{hines07} Hines, D.~C., Schneider, G., Hollenbach, D., et~al. 2007, \apjl, 671, L165 
\bibitem[2008]{kalas08} Kalas, P., et al. 2008, Science ~322, 1345 
  \bibitem[2010]{liseau} Liseau,  R., et al. 2010, this  volume 
  \bibitem[2010]{matthews} Matthews et al. 2010, this  volume 
  \bibitem[2010]{pilbratt} Pilbratt, G., et al. 2010, this volume
  \bibitem[2010]{poglitsch} Poglitsch, A., et al. 2010, this volume
\bibitem[2004]{stapelfeldt04} Stapelfeldt, K.R., et al. 2004, \apjs ~154, 458 
\bibitem[1996]{stern} Stern, S.A. 1996, \aap ~310, 999
\bibitem[2006]{su} Su, K.Y.L., Rieke, G.H., Stansberry,J.A. et al. 2006, \apj ~653, 675
\bibitem[2008]{trilling08} Trilling et al. 2008, \apj ~674, 1086
\bibitem[1999]{wyatt99} Wyatt, M.C., et al. 1999, \apj ~527, 918 
 \end{thebibliography}
\end{document}